\documentclass[12pt]{article}
\usepackage{amsfonts}

\newcommand{\eq}{\begin{equation}}
\newcommand{\eqx}{\end{equation}}
\newcommand{\eqn}{\begin{eqnarray}}
\newcommand{\eqnx}{\end{eqnarray}}
\newcommand{\f}[2]{\frac{#1}{#2}}

\newcommand{\cor}[1]{\left\langle{#1}\right\rangle}

\newcommand{\chiu}{\chi^{U(1)}}
\newcommand{\chiv}{\chi^{Vir}}
\newcommand{\ishi}[1]{|{#1}\rangle\!\rangle}
\newcommand{\ket}[1]{|{#1}\rangle}
\newcommand{\al}{\alpha}
\newcommand{\bt}{\beta}
\renewcommand{\th}{\theta}
\newcommand{\qt}{\tilde{q}}
\newcommand{\nn}{{\cal N}}

\newcommand{\arrb}[4]{%
\left[\begin{tabular}{cc}
$#1$ & $#2$ \\
$#3$ & $#4$
\end{tabular}\right]}

\newcommand{\args}{\arrb{1}{1}{1}{1}}
\newcommand{\FF}{{\cal F}}
\newcommand{\qq}{\quad\quad}
\newcommand{\LL}{{\cal L}}
\newcommand{\ZZ}{{\mathbb{Z}}}

\newcommand{\woz}{/\{0\}}

\title{Exceptional boundary states at $c=1$}

\author{Romuald A. Janik\footnote{{\tt
e-mail:ufrjanik@if.uj.edu.pl}}\\ \\
M. Smoluchowski Institute of Physics\\ 
Jagellonian University\\ 
Reymonta 4, 30-059 Cracow, Poland}

\begin{document}

\maketitle

\abstract{We consider the CFT of a free boson compactified on a
circle, such that the compactification radius $R$ is
an irrational multiple of $R_{selfdual}$. Apart from the 
standard Dirichlet and Neumann boundary
states, Friedan suggested \cite{FRIEDAN} that an additional
1-parameter family of
boundary states exists. These states break $U(1)$ symmetry of the
theory, but still preserve conformal invariance. 
In this paper we give an explicit construction
of these states, show that they are uniquely determined by the
Cardy-Lewellen sewing constraints, and we study the spectrum in the
`open string channel', which is given here by a continous integral with a
nonnegative measure on the space of conformal weights.}

\section{Introduction}

Boundary states in Conformal Field Theories (CFT) correspond to
boundary conditions that one may impose on the CFT that do not destroy
conformal invariance. The study of these states (or equivalently 
boundary conditions)
is interesting both from the point of view of statistical mechanics,
where one can gain a lot of information on the behaviour of
two-dimensional critical systems in confined geometries \cite{BSSTAT},
and from the point of view of string theory where these states
correspond to D-branes, nonperturbative objects in string theories
\cite{GEPNER,DV}. 

The theoretical description of these states and, more generally,
boundary conformal field theory (BCFT), is very well developed in the
context of rational theories (i.e. with a finite number of primary
fields w.r.t Virasoro or an extended chiral algebra)
\cite{CARDY,CL,L,SF,ZUBER}. However once we relax the requirement of
rationality (by e.g. breaking the invariance w.r.t. an extended chiral
algebra) new phenomena may appear such as moduli spaces
for boundary states. In addition, continuous spectra of conformal
weights may make the meaning of Cardy's condition for boundary states 
unclear.
It is thus very interesting to investigate BCFT in the 
nonrational case.

In this work we would like to consider boundary states in a
(compactified) free boson theory. This theory has an affine $U(1)$
symmetry, and the boundary states preserving that are the classical
Neumann and Dirichlet b.s. When the boundary states break $U(1)$ but
still preserve conformal symmetry, Friedan \cite{FRIEDAN}
suggested that a continous family of boundary states should appear
parametrized by a real interval\footnote{This is the case when the
compactification radius is such that $R/R_{selfdual}$ is irrational}. 
The goal of this paper is to propose
an explicit construction of these states and to study some of their
properties like the `open string channel' spectrum.

The interest in such symmetry breaking conditions lies also in the
fact that usually a lot of symmetry which makes a theory rational
leads to just a finite number of boundary states, like e.g. a finite
number of D-brane states in the Gepner models of Calabi-Yau
compactifications \cite{GEPNER}. In order to explore D-branes with
nontrivial moduli space in this context one has to impose much less stringent
restrictions on the symmetry properties of the states.

The plan of this paper is as follows. In section 2 we will briefly
recall basic properties of the $c=1$ discrete states and some features
of BCFT. In section 3 we state the proposal for the explicit form of
Friedan's boundary states. Then we show that the coefficients of these
boundary states are in fact uniquely determined by the Cardy-Lewellen
sewing conditions. In section 4 we show it for the lowest nontrivial
case, while the general, more technical proof is relegated to the appendix.
In section 6 we make the transformation into the open string channel. 
We close the paper with a discussion and a summary.

\section{$c=1$ discrete states and BCFT}

Let us consider a free boson compactified on a circle of
radius $R$. The partition function of this theory is
\eq
Z=\f{1}{\eta(\tau)\bar{\eta}(\bar{\tau})} \sum_{e,m \in \ZZ}
q^{(e/R+mR/2)^2/2} \bar{q}^{(e/R-mR/2)^2/2} 
\eqx
This CFT has $c=1$ and posseses a $U(1)$ current algebra.
The terms in the sum are characters of
$U(1)$ Kac-Moody representations labelled by $U(1)$ charges. 

In this paper we will restrict ourself to the case when $R$ is an
irrational multiple of $R_{selfdual}$. 
Then all the $U(1)$ representations for $e\neq 0$
or $m\neq 0$ are also irreducible w.r.t the
Virasoro algebra.
Only for $e=m=0$ (the vacuum sector) the $U(1)$ representation decomposes
into an infinite sum of Virasoro representations associated with the
so-called discrete states \cite{KAC,SEGAL}:
\eq
\chiu_{e=m=0}(q)=\f{1}{\eta(q)} = \sum_{J=0}^\infty \chiv_{J^2}(q)
\eqx
where
\eq
\chiv_{J^2}(q)=\f{q^{J^2}-q^{(J+1)^2}}{\eta(q)}
\eqx  
The corresponding (unnormalized) Virasoro primary fields are given by the
expressions \cite{POLKLEB}:
\eq
\label{e.fields}
V_J=\left( \int \f{du}{2\pi} e^{-i \sqrt{2}X(u+z)} \right)^J e^{i\sqrt{2}JX(z)}
\eqx
The free boson $X(z)$ is normalized here so that the conformal weight
$h_J=J^2$.

Boundary states are constructed from Ishibashi states \cite{ISHI} 
labelled by Virasoro primaries. The standard boundary states
corresponding to Neumann and Dirichlet boundary conditions are given
by the expressions (see e.g. \cite{AO}):
\eqn
\label{e.neu}
\ket{N(\theta)}&=& \sum_{J=0}^\infty (-1)^J\ishi{J} +\sum_{m\in \ZZ\woz}
e^{im\theta} \ishi{(0,m)} \\
\label{e.dir}
\ket{D(\theta)}&=& \sum_{J=0}^\infty \ishi{J} +\sum_{e\in \ZZ\woz}
e^{ie\theta} \ishi{(e,0)}
\eqnx 
where $\ishi{(e,0)}$ (resp. $\ishi{(0,m)}$) correspond to vertex
operators with the same (resp. opposite) $U(1)$ charges in the
holomorphic and antiholomorphic sectors.
It is well known \cite{CK,PT,RS} that for special values of the
compactification radius additional boundary states may appear.
However even for the generic case of  {\em irrational} $R/R_{selfdual}$ 
Friedan suggested \cite{FRIEDAN} that a 1-parameter family of boundary
states should appear.
An explicit construction of this family is the
main aim of this paper.
Before we proceed let us first recall some features of BCFT.

In general we have an expansion
\eq
\label{e.bishi}
\ket{\al}=\sum_{J=0}^\infty B^\al_J \ishi{J}+\sum_{e\in \ZZ\woz} B^\al_e
\ishi{(e,0)} +\sum_{m \in \ZZ\woz} B^\al_m \ishi{(0,m)}
\eqx
The coefficients $B_{J,e,m}$ are required to satisfy a number of
constraints.

One fundamental
property that one has to impose is that the spectrum of the theory between
any two such states decomposes into characters of Virasoro
representations (Cardy's condition):
\eq
\label{e.cardy}
\sum_i B^\al_i B^\bt_i \chiv_{i}(q)=\sum_A n_{\al\bt}^A \chiv_A(\qt)
\eqx
where $n^A_{\al\bt}$ are some coefficients, $q=e^{2\pi i \tau}$, $\qt$
is related to $q$ by the modular transformation $\tau \to -1/\tau$,
and $\chiv_{i,A}$ are Virasoro characters at the same central charge.

When we are in the
rational setting the sum is finite and the coefficients have to be
integers (this is the main content of Cardy's condition). 
In the nonrational case the summation is transformed into an
integral and the meaning of Cardy's condition then is not completely
understood, in particular it is not obvious what conditions one should 
impose on the $n^A_{\al\bt}$ which now is a `density function'
(measure) on the space of conformal weights $A$.

The least that we may do is to require that for {\em any} choice of
boundary states $\ket{\al}$, $\ket{\bt}$ the density function
$n^A_{\al\bt}$ should be {\em non-negative}. In section 6 we will
verify that this property holds for our construction of Friedan's
boundary states.

Apart from Cardy's condition, boundary state coefficients have to
satisfy a set of sewing constraints \cite{CL,L,ZUBER}, which link them
with other ingredients of a BCFT such as bulk-to-boundary operator
product coefficients, OPE between boundary fields, 1-point functions etc.

We will show that these constraints determine unambigously all the
coefficients of (\ref{e.bishi}).
From this point of view, the most interesting
quantities in the BCFT are the 1-point functions $A^\al_i$ defined through
\eq
\cor{\Phi_i(z,\bar{z})}_\alpha=\f{A^\al_i}{|z-\bar{z}|^{2h_i}}
\eqx
where the superscript $\alpha$ denotes that the average is performed on
the upper half plane with the boundary condition $\ket{\al}$ on the
real axis.
There is a very close relation of these data to the coefficients of
the boundary state (\ref{e.bishi}), first obtained in \cite{CL}:
\eq
A^\al_i=\f{B^\al_i}{B^\al_0}
\eqx
where $0$ denotes the identity. 
In the following we will normalize the coefficient of the identity to
be 1, which allows us to identify the coefficients $B^\al_i$ with the
1-point functions $A^\al_i$.

Cardy and Lewellen derived a nonlinear constraint on the 1-point
functions \cite{CL,L,ZUBER} which we will use in section 4 to
determine the form of the boundary states (\ref{e.bishi}).

\section{A proposal for Friedan's boundary states}

We propose the following expression for the boundary states
(\ref{e.bishi}) constructed
using the Ishibashi states corresponding to the fields
(\ref{e.fields}):
\eq
\label{e.prop}
\ket{x}=\nn \sum_{l=0}^\infty P_l(x)\ishi{l}
\eqx
where $P_l(x)$ are the Legendre polynomials and $x \in (-1,1)$. In
particular the $U(1)$ charged states will have coefficient zero.
The normalization constant $\nn$ is undetermined and in the rest of
the paper we will set it equal to 1. We will use the coefficient of the
$\ishi{1}$ state ($P_1(x)=x$) as a label for the state. 

This choice of coefficients has the most natural behaviour under modular
transformations to the open string channel, which was initially our
main motivation for (\ref{e.prop}).
In addition we will
show that all the coefficients of (\ref{e.prop}) are indeed determined
{\em uniquely} as a function of $x$ through the Cardy-Lewellen sewing
conditions. 

In the next section we will show it explicitly for the coefficient of
$\ishi{2}$. The full proof is summarized in Appendix A.

Recently exactly this expression was reobtained through a certain
limiting procedure \cite{RECKGW} from theories with the
compactification radius being $R=\f{N}{M}R_{selfdual}$.

\section{Cardy-Lewellen sewing condition}

\label{s.lc}

From the point of view of determining the coefficients of the boundary
states the most interesting condition is the sewing constraint for
1-point functions:
\eq
\label{e.cl}
B^\al_i B^\al_j =\sum_k C^k_{ij} B^\al_k F_{k0}\arrb{j}{j}{i}{i}
\eqx
From now on we will drop the superscript $\al$ in $B^\al_i$.
This equation has been (implicitly or explicitly) 
used in recent studies of boundary
states in Liouville and $SL(2,\mathbb{R})$ WZW theory \cite{TESCH,ZAM,KUT}.
The $C^k_{ij}$ are the OPE coefficients for the bulk fields, while the
fusing matrix
$F_{k0}\arrb{j}{j}{i}{i}$ relates conformal blocks of argument $\eta$ to
conformal blocks at $1-\eta$:
\eq
\FF^p_{ij,ij}(1-\eta)=\sum_q F_{pq}\arrb{j}{j}{i}{i} \FF^q_{jj,ii}(\eta) 
\eqx
Here the conformal blocks are normalized as $\FF^q_{il,jk}(\eta) \sim
1 \cdot \eta^{h_q-h_j-h_k}$ for small $\eta$.  

Let us now set $i=1$ (primary field of conformal weight
$h=1^2=1$) in the CL sewing equation (\ref{e.cl}).
Then it is easy to see that given the value of
\eq
B_1=x
\eqx
{\em all} the other coefficients are uniquely fixed. In order to
verify that the proposal (\ref{e.prop}) is correct we have to verify
that (\ref{e.cl}) becomes equivalent to the recursion relation for the
Legendre polynomials:
\eq
\label{e.legrec}
x P_j(x)=\f{j}{2j+1} P_{j-1}+\f{j+1}{2j+1}P_{j+1}
\eqx
We will now verify explicitly the lowest nontrivial case ($j=1$), and
state the general more technical argument in Appendix A.

From the explicit form of the primary fields (\ref{e.fields}) we see
that the OPE coefficient $C^k_{1j}$ vanishes unless $k=j-1$ or
$k=j+1$ (see also appendix A). Therefore (\ref{e.cl}) takes the form
of a two term recursion relation similar to (\ref{e.legrec}). It
remains to show that the coefficients coincide. 

\subsubsection*{OPE coefficients}

We now have to compute the OPE's. The fields should first be normalized
so that the coefficient of the 2-point function on the plane is unity.
The $C^k_{ij}$ can be calculated through 3-point functions, which
follow from calculating the correlation function of the exponentials
of the free field $X$ and then performing the integrals over the $u$'s
by residues at $u=0$ (cf (\ref{e.fields})). The antiholomorphic sector
contributes equally and multiplicatively since the relevant operators
are diagonal. The results for the normalisations are
\eq
\cor{V_{1}V_{1}V_{0}}=2^2 \qq
\cor{V_{2}V_{2}V_{0}}=24^2 \qq
\cor{V_{3}V_{3}V_{0}}=720^2
\eqx
Here we supressed the standard conformal prefactor. The normalised
operators are therefore $O_0=V_0$, $O_1=\f{1}{2} V_1$,
$O_2=\f{1}{24}V_2$ and $O_3=\f{1}{720} V_3$. The lowest nonvanishing OPE's are
\eq
\cor{V_1V_1V_2}=8^2  \qq \cor{V_1V_2V_3}=144^2
\eqx
Which lead to the OPE coefficients:
\eq
C^0_{11}=1 \qq 
C^2_{11}=\f{1}{4}\cdot\f{1}{24}\cdot 8^2=\f{2}{3}  \qq
C^3_{12}=\f{1}{2}\cdot\f{1}{24}\cdot \f{1}{720}\cdot 144^2=\f{3}{5}
\eqx
The CL equation (\ref{e.cl}) at $j=1$ can then be rewritten as:
\eq
\label{e.clj}
x B_1=1 \cdot F_{00}\args B_0 + \f{2}{3} \cdot F_{20}\args B_2
\eqx
while the recursion relation for the Legendre polynomials takes the
form $xP_1=\f{1}{3} P_0(x)+\f{2}{3}P_2(x)$. It remains to determine
the fusing matrices. 

\subsubsection*{Fusing matrices}

We first have to determine the conformal blocks $\FF^0_{11,11}(\eta)$,
$\FF^1_{11,11}(\eta)$ and $\FF^2_{11,11}(\eta)$. To this end we note that
there is a level-3 singular vector in the Virasoro representation 
with $h=1$ and $c=1$, namely
\eq
(L_{-3}-\f{2}{3}L_{-1}L_{-2}+\f{1}{6}L_{-1}^3)\ket{1}
\eqx   
This can be directly translated into a $3^{rd}$ order differential
equation for the conformal blocks (see Appendix B), which has the
following three linear independent solutions
\eqn
f_0(\eta)&=&\f{1}{\eta^2(1-\eta)^2} \sim
\f{1}{\eta^2}\left(1+2\eta+\ldots\right)\\ 
f_1(\eta)&=&\f{1-\f{3}{2}\eta+\eta^2}{\eta(1-\eta)^2} \sim
\f{1}{\eta}\left(1+\ldots\right)\\ 
f_2(\eta)&=&\f{\eta^2}{(1-\eta)^2} \sim \eta^2(1+\ldots)
\eqnx
From the behaviour of the conformal blocks $\FF^p_{11,11}(\eta)\sim
\eta^{h_p-2}(1+\ldots)$ we can identify
$\FF^2_{11,11}(\eta)=f_2(\eta)$. From Virasoro 
algebra one can show that $\FF^0_{11,11}(\eta)=\eta^{-2}(1+0\cdot
\eta+2\eta^2+\ldots)$. Therefore
\eq
\FF^0_{11,11}(\eta)=f_0(\eta)-2f_1(\eta)+\gamma f_2(\eta)
\eqx
where $\gamma$ is still unknown. From these considerations we may now
determine the appropriate fusing matrices. Because
$f_2(1-\eta)=f_0(\eta)-4f_1(\eta)+f_2(\eta)$ and the coefficient of $f_0$ in
$\FF^0_{11,11}$ is 1, we may read off
\eq
F_{20}\args=1
\eqx
Analogously from $f_0(1-\eta)=f_0(\eta)$ and
$f_1(1-\eta)=\f{1}{2}f_0(\eta)-f_1(\eta)$ we get
\eq
F_{00}\args=\gamma
\eqx

It remains therefore to identify $\gamma$. In principle it could be
done by calculating the coefficient of $\eta^2$ in $\FF^0_{11,11}(\eta)$
which follows just from conformal symmetry \cite{BPZ}, but since such a
calculation is very tedious we will determine $\gamma$
indirectly. 

Consider the 4-point function 
\eq
\lim_{z_1 \to \infty} z_1^2 \bar{z}_1^2 \cor{O_1(z_1)O_1(1)O_1(\eta)O_1(0)}
\eqx 
Since $O_1$ is proportional to
$\partial X \bar{\partial}X$ we may calculate the above correlation
function  using Wick
contractions and express it in terms of the $f_i(\eta)$ with the result
\eq
(\f{1}{\eta^2}+1+\f{1}{(1-\eta)^2})\cdot c.c \equiv
(f_0(\eta)-2f_1(\eta)+f_2(\eta))\cdot c.c
\eqx
where $c.c.$ stands for complex conjugate.
On the other hand we know how the same correlation function is
expressed in terms of conformal blocks\footnote{Recall that the theory
is not diagonal.}:
\eq
\left(\sum_{p=0}^2 \sqrt{C^p_{11}}\sqrt{C^p_{11}} 
\FF^p_{11,11}(\eta)\right) \cdot c.c. =
\left| \FF^0_{11,11}(\eta) +\f{2}{3} \FF^2_{11,11}(\eta) \right|^2
\eqx
Keeping in mind that $\FF^2_{11,11}=f_2$ and
$\FF^0_{11,11}=f_0-2f_1+\gamma f_2$ we read off that
$\gamma=\f{1}{3}$. Thus we have finally 
\eq
F_{00}\args=\f{1}{3}
\eqx 
Note that this is exactly the value to make the CL equation
(\ref{e.clj}) to be consistent with the the recurrence relation for
the Legendre orthogonal polynomials. From this we get that
$B_2=\f{1}{2}(3x^2-1) \equiv P_2(x)$. 

In Appendix A we show that the CL equation for $i=1$ and arbitrary $j$
coincides with the recursion relation for the Legendre polynomials
(\ref{e.legrec}), thus fixing uniquely all the coefficients of the
discrete states to the values given in the proposal (\ref{e.prop}).

\section{Remarks}

We will now show that in general the $U(1)$ charged states in
(\ref{e.bishi}) have vanishing coefficients.
The CL equation (\ref{e.cl}) for the fields $i=1$ and $j=(e,0)$ takes the form
\eq
\label{e.clcharged}
x B_e = C^e_{1e} B_e F_{e0}\arrb{e}{e}{1}{1}
\eqx
Here we used the fact that there is a {\em unique} Virasoro primary
with charge $(e,e)$ for the theory with irrational
$R/R_{selfdual}$. We see that $B_e$ may be nonzero only for a
single value of $x$. By comparision with the formulas
(\ref{e.neu}-\ref{e.dir}) we get that $x=1$. In an analogous manner we
get that the $B_m$ coefficient may be nonzero only for $x=-1$. 

Therefore in the case $x=\pm 1$ the boundary states (\ref{e.prop}) become
superpositions of Dirichlet and Neumann boundary states i.e.
\eq
\label{e.limits}
\ket{1}=\f{1}{2\pi} \int_0^{2\pi} d\th \ket{D(\th)} 
\quad\quad\quad
\ket{-1}=\f{1}{2\pi} \int_0^{2\pi} d\th \ket{N(\th)} 
\eqx
A similar obervation has been made in \cite{RECKGW} for theories with
$R=\f{N}{M}R_{selfdual}$. 

In this context we see that one cannot obtain the family of Friedan's
boundary states from the conventional (Dirichlet or Neumann) theory
through deformation by some marginal boundary
operator, i.e. by writing the partition function on the disk as
\eq
Z_{disk}=\int DX e^{-\f{1}{8\pi} \int_D d^2z \,\partial X
\bar{\partial}X +(x+1)\int_{\partial D} d\phi \,O_{boundary}(e^{i\phi})}
\eqx
The undeformed theory (with $x=-1$), as can be seen from
(\ref{e.limits}) should 
really be a superposition of theories with different boundary
conditions.
Therefore we cannot directly apply the methods of \cite{CK,PT,RS}
which were used to construct exceptional boundary states at special
values of the compactification radii.

In the remaining part of the paper we will calculate the spectrum of
the theory on a strip.  

\section{Open string channel spectrum}

Let us consider the partition function of the CFT on a cylinder on
the ends of which we impose boundary conditions $\ket{\cos \th_1}$ and
$\ket{\cos \th_2}$. In the `closed string channel' we have
\eq
\label{e.part}
Z=\sum_{l=0}^\infty P_l(\cos\th_1) P_l(\cos \th_2) \chiv_{l^2}(q)
\eqx
We will now show that in the `open string channel' this partition
function can be written as a continous integral of $c=1$ characters
$\chiv_h(\qt)$ (see (\ref{e.cardy})) with a {\em nonnegative} measure on the
space of conformal weights $h$.

We start with the formula
\eq
P_l(\cos \th_1)P_l(\cos \th_2)=\f{1}{\pi} \int_0^\pi P_l(\underbrace{\cos \th_1
\cos \th_2 -\sin \th_1 \sin \th_2 \cos \phi_2}_{\mbox{$\cos\th$}}) d\phi_2
\eqx
This expresses the product of two Legendre polynomials in terms of a
single one. Now it is convenient to express the resulting Legendre polynomial 
as an integral over the SU(2) group character through
\eq
P_l(\cos \th)=\f{1}{\pi} \int_0^{\pi} \f{\sin \left( l+\f{1}{2}
\right)t}{\sin \f{t}{2}} d\phi
\eqx
where $\cos t/2=\cos \th/2 \cos \phi/2$. It is this expression which
facilitates performing the modular transformation since 
\eq
\sum_{l=0}^\infty \f{\sin \left( l+\f{1}{2} \right)t}{\sin \f{t}{2}}
\left( q^{l^2}-q^{(l+1)^2} \right) =
1+2\sum_{n=1}^\infty q^{n^2} \cos nt
\eqx
Now we can use Poisson resummation to easily perform the modular
transformation to obtain
\eq
\f{1}{\sqrt{-i2\tau}} \sum_{n=-\infty}^{+\infty}
\qt^{\f{1}{4}\left(n+\f{t}{2\pi} \right)^2}
\eqx
The $\sqrt{-i\tau}$ will be compensated by the transformation formula
for the Dedkind eta function: $\eta(-1/\tau)=\sqrt{-i\tau} \eta(\tau)$.

Finally we obtain for the partition function (\ref{e.part}) but now
viewed in the `open string channel': 
\eq
\label{e.rhs}
Z=\f{1}{\sqrt{2}\pi^2} \int_0^\pi d\phi_2 \int_0^{\pi} d\phi 
\sum_{n=-\infty}^{+\infty} \f{\qt^{\f{1}{4}\left(n+\f{t}{2\pi}
\right)^2}}{\eta(-1/\tau)} 
\eqx
where $t$ depends on $\th_1,\th_2,\phi$ and $\phi_2$ through 
\eqn
\cos \f{t}{2} &=&\cos \f{\th}{2} \cos \f{\phi}{2}\\
\cos \th &=& \cos \th_1 \cos \th_2 -\sin \th_1 \sin \th_2 \cos \phi_2
\eqnx
In the above formulas we thus get a combination of Virasoro/$U(1)$ characters:
\eq
Z=\f{1}{\sqrt{2}\pi^2} \int_0^\pi d\phi_2 \int_0^{\pi} d\phi
\sum_{n=-\infty}^{+\infty} \chi_{\f{1}{4}\left(n+t/2\pi\right)^2}(\qt)
\eqx
The conformal weights $h=\f{1}{4}(n+t/2\pi)^2$ appear in bands of finite width
parametrized by the angles $\phi$ and $\phi_2$:
\eq
t(\th_1,\th_2,\phi,\phi_2)=2\arccos \left\{ \cos \f{\phi}{2} 
\sqrt{ \f{1+ \cos \th_1 \cos \th_2 -\sin \th_1 \sin \th_2 \cos \phi_2}{2}}
\right\}
\eqx
We see that the measure is explicitly nonnegative for any choice of
$\theta_1$ and $\theta_2$. This requirement limits, in a natural way,
the range of the parameter $x$ labelling the boundary states $\ket{x}$
(\ref{e.prop}) to lie between -1 and 1. If $x$ would be outside this
interval, complex conformal weights would be generated.

\section{Discussion}

In this paper we constructed explicitly the 1-parameter family of
boundary states, mentioned in \cite{FRIEDAN}, in a $c=1$ free boson 
compactified on a circle of
radius $R$ with $R/R_{selfdual}$ irrational. We showed, in
particular, that the coefficients of these boundary states are
uniquely fixed by the Cardy-Lewellen sewing conditions. This implies
that at these radii no additional boundary states may appear.
Once we move away frome these radii, additional $U(1)$ charged states
may enter the expression for boundary states and there may appear
additional parameters. Technically this arises due to the appearance
of additional terms on the rhs of (\ref{e.clcharged}). This
complementary case of $R/R_{selfdual}$ {\em rational} is studied
in \cite{RECKGW}.

An interesting feature of the family of Friedan's boundary states is
that they may not be reached by a deformation of a single standard
(i.e. Neumann or Dirichlet) theory by a marginal boundary operator.

This gives rise to an intriguing open question what is the Lagrangian
formulation of a free boson theory with Friedan's boundary conditions.
In this paper these boundary conditions were determined algebraically
by fixing the coefficients of appropriate Ishibashi states. A direct
understanding, however, of the meaning of these conditions in terms of
the field $X(z,\bar{z})$ would be very interesting.

The answer to these questions might also shed some light on another
puzzling feature of these states, namely the continous spectrum of
conformal weights in the open string channel. Cardy's condition is
now much less restrictive than in the rational case. The
multiplicities are no longer constrained to be positive integers, 
but become a measure
on the space of conformal weights. It seems that the only requirement
that we may impose is just the positivity of the measure.
And indeed we checked that for any two states from this family the relevant
measure turns out to be {\em nonnegative}.

The structure of the space of boundary states even for the simplest
example of an irrational free boson CFT turns out to be quite
complex. The irrational BCFT's exhibit several new phenomena like
continous spectra and  nontrivial structure of the moduli spaces of states.
It would be very interesting, both by its own right and in view of
potential applications to D-branes, to understand these new features
more systematically for a wider class of irrational BCFT's.     

\subsubsection*{Acknowledgements}

I would like to thank Mathias Gaberdiel and Andreas Recknagel for
discussions of their related work, Jean-Bernard Zuber for providing a
copy of \cite{FRIEDAN}, for discussions, and for encouragement for 
writing up the results of this investigation. 
This work was supported in part by KBN grant 2P03B01917.


\section*{Appendix A --- recursion relations (general case)}

Here we will show that the CL sewing conditions for any $k$
\eq
\label{e.recgen}
x B_k = F_{k-1\,0}\arrb{k}{k}{1}{1} C^{k-1}_{1k} B_{k-1}+
F_{k+1\,0}\arrb{k}{k}{1}{1} C^{k+1}_{1k} B_{k+1}
\eqx
coincides with the recursion relation for the Legendre polynomials,
thus determining uniquely the coefficients of the boundary state
(\ref{e.prop}).

\subsubsection*{Step 1}

We will first show that
\eq
\label{e.fus}
F_{k+1\,0}\arrb{k}{k}{1}{1}=1
\eqx
This fusing matrix is defined through the crossing property of
conformal blocks:
\eq
\FF^{k+1}_{1k,1k}(1-\eta)=\sum_l F_{k+1\,l}\arrb{k}{k}{1}{1}
\FF^l_{kk,11}(\eta)
\eqx
We note that the identity $l=0$ has the most singular behaviour $\sim
\eta^{-2}$ on the right hand side, so we can isolate the relevant
fusing matrix through
\eq
\label{e.lim}
F_{k+1\,0}\arrb{k}{k}{1}{1}=\lim_{\eta \to 0} \eta^2
\,\FF^{k+1}_{1k,1k}(1-\eta)
\eqx
The conformal block $\FF^{k+1}_{1k,1k}(\eta)$ can be determined from the
appropriate differential equation (Appendix B). As in section
\ref{s.lc} there is no ambiguity, as this conformal block behaves
like $\eta^{2k}$:
\eq
\label{e.cbk}
\FF^{k+1}_{1k,1k}(\eta)=\f{\eta^{2k}}{(1-\eta)^2}
\eqx
Inserting (\ref{e.cbk}) into (\ref{e.lim}) we get (\ref{e.fus}).

\subsubsection*{Step 2}

We will show that the OPE coefficient $C^{k+1}_{1k}$ is
\eq
\label{e.opegen}
C^{k+1}_{1k}=\f{k+1}{2k+1}
\eqx
It is easy to see that the OPE coefficients between the discrete
states $O_J$ do not depend on the radius of compactification.
Therefore for the computation, we may choose the self dual
radius. Then there appear additional Virasoro primaries $O_{J,m}$,
which form multiplets of $SU(2)$
\eq
O_{J,m}(0) \propto \left( :\int \f{du}{2\pi i} e^{-i\sqrt{2}X(u)}:
\right)^{J-m} \!\!\!\!\! :e^{i\sqrt{2}JX(0)}: =(J_0^-)^{J-m} 
:e^{i\sqrt{2}JX(0)}:
\eqx
Using $SU(2)$ symmetry, as was argued in \cite{POLKLEB}, the OPE
coefficients (just for the holomorphic parts) have the form
\eq
\label{e.cgope}
O_{J_1,m_1} O_{J_2,m_2}\sim \sum_{J,m}
C(\{J_1,m_1\},\{J_2,m_2\},\{J,m\}) f(J_1,J_2,J) O_{J,m}
\eqx
where $C(\{J_1,m_1\},\{J_2,m_2\},\{J,m\})$ is the $SU(2)$
Clebsch-Gordan coefficient and $f(J_1,J_2,J)$ is some unknown function which
remains to be determined\footnote{This function is different from the
one determined in  \cite{POLKLEB}, since there the $O_{J,m}$ have been
`gravitationally dressed'.}. 
We can fix the function $f(1,k,k+1)$ by calculating directly
\eq
\label{e.fnorm}
O_{1,1}O_{k,k}\sim 1\cdot O_{k+1,k+1}
\eqx
Here we used the fact that the $O_{J,J}$ operators are just equal to
the vertex operators. The normalization of the $O_{J,m}$ operators
follow from $SU(2)$ symmetry. 
Therefore using (\ref{e.cgope}) and (\ref{e.fnorm}) we have 
\eq
C^{k+1}_{1k}=\left[ \f{C(\{1,0\},\{k,0\},\{k+1,0\})}{ 
C(\{1,1\},\{k,k\},\{k+1,k+1\})} \right]^2=\f{k+1}{2k+1} 
\eqx
which is the result we wanted to obtain.

\subsubsection*{Step 3}

It remains to determine the remaining fusing matrix. Again for the same
reason as discussed in section \ref{s.lc} it is difficult to do it
directly. Therefore we will use crossing symmetry for (the holomorphic
part of) the correlation function $\cor{O_1O_kO_kO_1}$:
\eq
\sum_p \sqrt{C^p_{1k} C^p_{1k}} \FF^p_{1k,1k}(\eta)= 
\sum_q \sqrt{C^q_{11} C^q_{kk}} \FF^q_{kk,11}(1-\eta)
\eqx
We may now reexpress the conformal blocks $\FF^p_{1k,1k}(\eta)$ in
terms of $\FF^q_{kk,11}(1-\eta)$ using fusing matrices, and equate the
coefficients of $\FF^0_{kk,11}(1-\eta)$. In this way we obtain
\eq
\sum_p C^{p}_{1k} F_{p0}\arrb{k}{k}{1}{1}=\sqrt{C^0_{11} C^0_{kk}}=1
\eqx
where the sum runs only over $p=k-1$ and $p=k+1$. 
From here we get the result
\eq
\label{e.fustwo}
F_{k-1\, 0}\arrb{k}{k}{1}{1} = \f{1-C^{k+1}_{1k}}{C^{k-1}_{1k}} =
\f{1-C^{k+1}_{1k}}{C^{k}_{1\,k-1}}=
\f{1-\f{k+1}{2k+1}}{\f{k}{2k-1}}=\f{2k-1}{2k+1}
\eqx

\subsubsection*{Final result}

When we insert the results (\ref{e.fus}), (\ref{e.opegen}) and
(\ref{e.fustwo}) into (\ref{e.recgen}), and use $C^{k-1}_{1k} \equiv
C^k_{1\, k-1}=\f{k}{2k-1}$, 
we obtain the recursion relation for the coefficients $B_k$:
\eq
x B_k=\f{k}{2k+1} B_{k-1}+\f{k+1}{2k+1} B_{k+1}
\eqx
This allows us to unambigously identify $B_k=P_k(x)$.

\section*{Appendix B --- differential equations for conformal blocks}

The differential equation for the conformal blocks $\FF^p_{11,11}$ is
obtained by writing the differential equation for the 4-point
correlation function:
\eq
(\LL_{-3}-\f{2}{3}\LL_{-1}\LL_{-2}+\f{1}{6}\LL_{-1}^3)
\underbrace{\left[\prod_{i<j} (z_i-z_j)^{-\f{2}{3}} G(x)\right]}_
{ \cor{O_1(z_0)O_1(z_1)O_1(z_2)O_1(z_3)}} =0
\eqx
and then taking the limit $z_1 \to 0$, $z_2 \to 1$, $z_3 \to \infty$
and $z_0\to \eta$. When we substitute $G(x)=\eta^{\f{2}{3}} (1-\eta)^{\f{2}{3}}
\FF(\eta)$ we obtain the following equation for the conformal blocks
$\FF^p_{11,11}(\eta)$:
\eqn
&&(4-9\eta)\FF+2\eta(\eta-1)(5\eta^2-5\eta-1)\FF'+ \nonumber\\
&&  +4\eta^2(\eta-1)^2(2\eta-1)\FF'' +\eta^3(\eta-1)^3 
\FF'''=0
\eqnx

\noindent{}The relevant equation for conformal blocks of
$\cor{O_1(z_0)O_k(z_1)O_1(z_2)O_k(z_3)}$ used in Appendix A is
\eqn
&& -4(k^2 (\eta-1)^3-\eta(1-3\eta+\eta^2))\FF -
2(\eta-1)\eta(-1+2k^2(\eta-1)^2+\nonumber\\
&& +9\eta-7\eta^2) \FF'+ 
 4(\eta-1)^2\eta^2 (2\eta-1) \FF'' +\eta^3(\eta-1)^3
\FF'''=0
\eqnx
The linear independent solutions of this equation are
\eq
\f{\eta^{-2k}}{(1-\eta)^2}\ , \quad\quad\quad 
\f{2k^2- 4k^2 \eta+\eta+2k^2 \eta^2}{\eta k^2(1-\eta)^2}\ , \quad\quad\quad
\f{\eta^{2k}}{(1-\eta)^2}
\eqx
The solution with the leading coefficient $\eta^{2k}$ may be
unambigously identified with the conformal block
$\FF^{k+1}_{1k,1k}(\eta)$.

\pagebreak

\end{document}